\documentclass[aps,twocolumn,showpacs,amsmath,amssymb]{revtex4}
\usepackage{dcolumn}
\usepackage{bm}
\usepackage{color}
\usepackage{latexsym}
\usepackage{amssymb}
\usepackage{graphicx}
\usepackage{ulem}

\begin{document}
\title{On the magnetotransport of 3D systems in quantizing magnetic field}
\author{M. V. Cheremisin}
\affiliation{A.F.Ioffe Physical-Technical Institute,
St.Petersburg, Russia}
\date{\today}
\begin{abstract}
The resistivity components of 3D electron gas placed
in quantizing magnetic field are calculated taking into account the correction
caused by combined action of the Peltier and Seebeck thermoelectric effects. The longitudinal,
transverse and the Hall magnetoresistivities exhibit familiar 1/H-period
oscillations being universal functions of magnetic field and temperature.
\end{abstract}


\maketitle

The electronic transport of 3D solids subjected to strong magnetic
field has been intensively studied since the observation of the
Schubnikov-de Haas oscillations[1]. The most of the interest
concerned the transverse (i.e $\bot B$) magnetotransport. The problem was
considered semiclassically[2,3] and,
then, using the rigorous quantum mechanical approach[4,5].
With the help of density matrix equation the components of the conductivity
tensor associated with transverse magnetotransport was obtained
in Ref.[4]. Despite this progress, the incorporation of the thermal
effects faced the strong difficulties[6-9].
As it was demonstrated for the first time in Ref.[10], in
quantizing magnetic field it is necessary to take into account
the diamagnetic surface currents. Neglecting dissipation the transverse
thermal current and energy flux are given by the sum of bulk and surface
components. With the diamagnetic currents accounted[10] the
transport coefficients satisfies the phenomenological Einstein and Onsager
relationships.

\section{General formalism}
\label{General formalism}
The main goal of the present paper concerns the peculiar
thermodynamic approach[11-13]
regarding 3D electron transport in strong magnetic fields. We will
follow the argumentation first put forward by Kirby and Laubitz [11], and then
modified in Refs.[12,13] for magnetotransport in
2D electron(hole) systems. This approach allows one to account for both the
Schubnikov-de Haas Oscillations and Integer Quantum Hall Effect[14] modes.
Let us consider the isotropic sample subjected
in strong magnetic field $B=B_{z}$. Neglecting spin
splitting, the 3D energy spectrum yields[15]
\begin{equation}
\varepsilon _{n}=\hbar
\omega _{c}(n+1/2)+\varepsilon_{\|},
\label{3D_spectrum}
\end{equation}
where $n=0,1$... is the LL number, $\omega _{c}=eB/mc$ is the cyclotron
frequency, $m$ is the electron effective mass. Then, $p_{z}$ and $\varepsilon
_{\|}=\frac{p_{z}^{2}}{2m}$ are the momentum and the kinetic energy respectively of
an electron moving along the magnetic field. The electron motion in x-y plane
is quantized, thus results in discrete Landau level set. For simplicity, we further
neglect the broadening of the Landau levels. The density of states associated
with the certain LL obeys the textbook result $\Gamma=1{\bf /}2\pi l_{B}^{2}$,
where $l_{B}=(\hbar c/eB)^{1/2}$ is the magnetic length. In the present paper
we mostly interest in the strong quantum limit case, when $\hbar \omega _{c}>>kT$.

In general, the macroscopic current, ${\bf j}$
,\hspace{0in} and the energy flux, ${\bf q}$ , densities are given by[16]
\begin{eqnarray}
{\bf j}=\hat{\sigma}({\bf E-}\alpha {\bf \nabla}T),
\label{current+heat} \\
{\bf q}=\left( \alpha T-\zeta /e\right) {\bf j}-{\bf \hat{\kappa} }{\bf \nabla }T.
\nonumber
\end{eqnarray}
Here, ${\bf E}=\nabla \zeta /e$ is the electric field, $\mu$ and $\zeta =\mu
-e\varphi
$ are the chemical and electrochemical potential respectively, $\alpha $ is the thermopower. Then, $\hat{\sigma}$ and $\hat{\kappa}=LT\hat{\sigma}$ are the conductivity and thermal
conductivity tensors respectively, $L=\frac{\pi^{2}k^{2}}{3e^{2}}$ is the Lorentz number. It will be recalled
that Eq.(\ref{current+heat}) is valid for a confined-topology sample for which the diamagnetic surface
currents [10] are taken into account. Both the Einstein and
Onsager relationships are satisfied. The movement of the electron along the
magnetic field is not quantized, thus results in Drude longitudinal component $\sigma _{zz}=Ne^{2}\tau /m$ of the conductivity tensor, where $N$ is the 3D electron density. In what follows we assume the constant momentum relaxation time $\tau$ in z-direction. Compared to unperturbed movement of an electron along the magnetic field, the transverse drift of electrons in x-y plane results in off-diagonal Hall components[4] $\sigma_{yx}=-\sigma_{xy}=Nec/B$ of the conductivity tensor. Then, in contrast to previous studies[2-5] we suggest that in strong quantum limit $\sigma_{xx}=\sigma_{yy}=0$. Actually, the x-y plane related fraction of Eq.(\ref{current+heat}) can be viewed[7] as the current and energy dissipationless fluxes caused by electron drift in crossed fields. With the help of the above notations, the resistivity tensor $\hat{\rho}=\hat{\sigma }^{-1}$ obeys the same symmetry, namely $\rho_{yx}=-\rho_{xy}=1/\sigma_{yx}$, $\rho_{zz}=1/\sigma_{zz}$. The crucial idea of the present paper is that a \textbf{nonzero transverse magnetoresistivity} $\rho$ can, nethereless, arise due to combined action of the Peltier and Seebeck thermoelectric effects[11-13].

\section{Results and discussion}
\label{Results and discussion}

\begin{figure}[tbp]
\begin{center}\leavevmode
\includegraphics[width=0.8\linewidth]{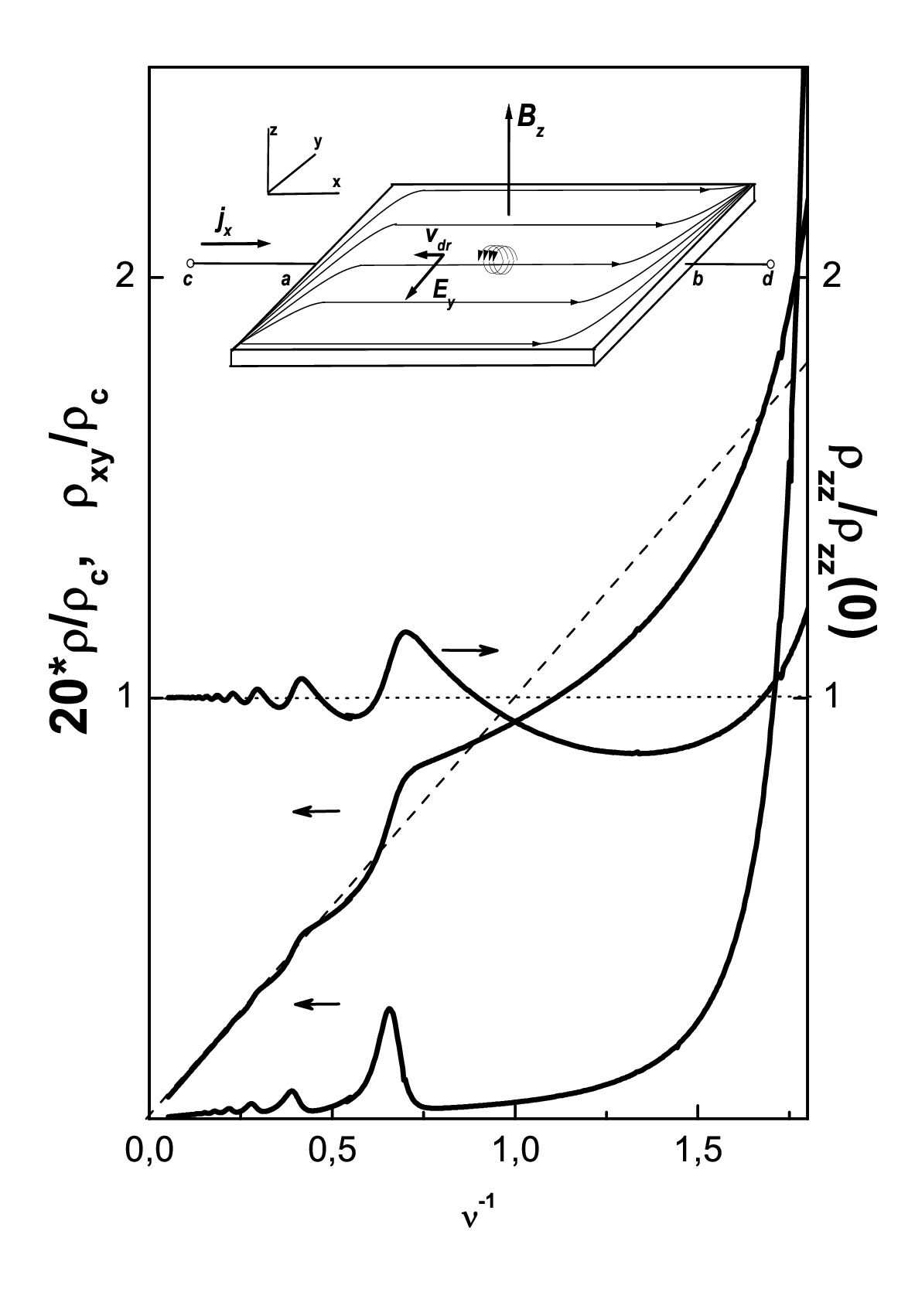}
\caption{\label{fig:Fig.1} Dimensionless Hall resistivity $\rho
_{yx}$, transverse magnetoresistivity $\rho$( both scaled by $\rho _{c}$) and the longitudinal resistivity $\rho _{zz}/\rho
_{zz}(0)$ vs magnetic field $B \sim \nu ^{-1}$ for certain value of the dimensionless temperature $\xi =0.025$. The
dashed(dotted) line represents the classical result for Hall and longitudinal resistivity respectively. Inset: the experimental
setup.}
\end{center}
\end{figure}

Let us discuss in details the typical experimental setup allowed to measure the Hall resistivity $\rho_{yx}$ and transverse magnetoresistivity $\rho$. The standard Hall-bar geometry sample (Fig.\ref{fig:Fig.1}, inset) is connected to the current source
by means of two identical leads. The contacts "a" and "b" are ohmic. The longitudinal voltage is measured
between the open ends ("e" and "d"), maintained
at the ambient temperature. The sample is placed in a chamber ( not shown in Fig.\ref{fig:Fig.1}, inset) kept at the bath
temperature $T_{0}$. Note that the electron temperature may, in principle, exceed the bath temperature if the
electron-phonon coupling is weak at low temperatures. Actually, the 3D electron gas cooling
can occur predominantly through the contacts of the sample and the
leads connected to them. However, the heat leakage via metal leads can be disregarded similar
to that reported[17] for two-dimensional electron case. Finally, we will consider the
adiabatic cooling of 3D electrons.

Recall that the Peltier heat is generated by a current across the interface between two
different conductors. At the contact ( e.g. "a" in
Fig.\ref{fig:Fig.1}, inset), the temperature, $T_{a}$,
electrochemical potential $\zeta $, normal components of the total
current, $I$, and the total energy flux are continuous. There
exists a difference $\Delta \alpha =\alpha _{m}-\alpha $ between
the thermoelectric powers of the metal conductor and 3D sample,
respectively. Then, $Q_{a}=I\Delta \alpha T_{a}$ is the amount of
Peltier heat released per unit time in contact "a". For $\Delta
\alpha >0$ and current flow direction shown in
Fig.\ref{fig:Fig.1}, the contact "a" is heated and contact "b" is
cooled. The contacts are at different temperatures, and $\Delta
T=T_{a}-T_{b}>0$. At small currents, the temperature gradient is small and $%
T_{a,b}\approx T_{0}$. In this case the 3D sample thermopower $\alpha $
is nearly constant, hence, one can disregard the Thompson heating $\sim
IT\nabla \alpha $ in 3D bulk. Note that the amount of the Peltier heat
evolved at the contact "a" is equal to that absorbed at the contact "b".
The energy flux is continuous at each contact, thus Eq.(\ref{current+heat}) yields $\kappa
_{yx}\left. \nabla _{x}T\right| _{a,b}=-j\Delta \alpha T_{a,b}$. Here, we take into account that the current
is known[18] to enter and leave the sample at two diagonally
opposite corners ( Fig.\ref{fig:Fig.1}, inset). One can fulminantly find the longitudinal temperature gradient
$\nabla_{x}T=-\frac{j\Delta \alpha}{L \sigma _{yx}}$,
which is linear in current. Omitting the contribution of the conductor resistances, the voltage drop,
$U$, measured between the ends "e" and "d" is equal to Seebeck thermoelectromotive force
$U=\int E_{x}dx=\int\alpha dT=\Delta \alpha
(T_{a}-T_{b})$, thus gives the transverse magnetoresistivity $\rho =U/jl=(\Delta \alpha)^{2}\rho _{yx}/L$.

We now intend to find the actually measured longitudinal resistivity taking into account the
Peltier-Seebeck thermoelectric effects. Since the longitudinal movement of an electron unaffected by
magnetic field, we make use of the result[11] valid for $B=0$. The longitudinal resistivity of 3D electron gas acquire the correction $\Delta \rho_{zz}=(\Delta \alpha)^{2}\rho _{zz}/L$. Finally, we summarize the actually measured values of the 3D electron gas resistivities as it follows
\begin{equation}
\rho _{yx}=\frac{B}{Nec},\qquad \rho =s\rho _{yx},\qquad \rho _{zz}=(1+s)\frac{%
\rho _{yx}}{\omega _{c}\tau },
\label{resistivity}
\end{equation}
where we take into account that for the actual case of the metal leads $s=(\Delta \alpha)^{2}/L \simeq \alpha^{2}/L$.
Remind that in the strong quantum limit, the thermopower of dissipationless 3D
electron gas is a universal thermodynamic quantity proportional to
the entropy per electron[10]:
\begin{equation}
\alpha =-\frac{S}{eN},
\label{thermopower}
\end{equation}
where $S=-\left( \frac{\partial \Omega }{\partial T}\right) _{\mu
,B}$ is the 3D electron entropy, $N=-\left( \frac{\partial \Omega }{\partial \mu }\right)_{T,B}$ is the 3D electron density, $\Omega =-2kT\cdot \Gamma \sum\limits_{n,p_{z}}\ln(1+\exp ((\mu -\varepsilon _{n})/kT))$ is the thermodynamic potential, which accounts the spin degeneracy.

Ii is instructive to discuss the applicability of the Gibbs statistics formalism for actual 3D electrons case.
The make use of thermodynamic potential presumes a variable number of 3D particles while
the chemical potential is assumed to be a constant. We argue that this method is not prohibited since the chemical
potential of 3D electrons can be, for example, pined by an external
reservoir of carriers. The metal leads can play the role of the reservoir. Indeed, in
thermodynamic equilibrium the chemical potential of the multiple system
"3D electrons-conductor" flattens out. Let us suppose a certain value of the chemical
potential $\mu$, and, thus 3D density $N_{0}$ at $B=0$. The magnetic field growth may result
in subsequent changes of 3D electron density $N$. In strong fields the only few Landau levels are filled,
hence, the absolute change in $N$ can be of the order of the zero-field carrier density $N_{0}$.
Simultaneously, the number electrons in the external reservoir changes. The relative variation of the chemical
potential of the external reservoir yields $\delta \mu /\mu
=\frac{2}{3}N_{0}/N_{m} \sim 10^{-7}$, where $N_{m}\approx 10^{23}$cm$^{-3}$
is the electron density the metal leads, $N_{0}=10^{16}$cm$^{-3}$ is the typical density of 3D
solid. We conclude that the chemical potential of metal leads, and, hence 3D electrons
remains constant. Note that conventional reasoning of 3D electron plasma neutrality are rather
doubtful since the Debye screening length can be of the order of
macroscopic sample length scale( see Appendix \ref{Appendix} ) in quantizing magnetic field.

It can be demonstrated that both
the electron density $N$ and the entropy $S$ are universal
functions of the reduced temperature $\xi =kT/\mu $ and
dimensionless magnetic field $\hbar \omega _{c}/\mu =\nu ^{-1}$,
where $\nu $ is the so-called filling factor. Following the
conventional Lifshitz-Kosevich formalism[19], we can easily derive asymptotic equations for $N,S$,
and, hence, for $\alpha $, valid in the low-temperature and
low-field limit $\nu ^{-1},\xi <1$:
\begin{eqnarray}
N &=&N_{0}\left( 1+\frac{\pi ^{2}}{2F_{1/2}\left( \frac{1}{\xi}\right)}\sum\limits_{k=1}^{%
\infty }\frac{(-1)^{k}\sin \left( 2\pi k\nu -\frac{\pi}{4}\right) }{\sqrt{z}\sinh z}%
\right) ,    \label{Lifshitz} \\
S &=&S_{0}+\frac{N_{0}k \pi ^{3}}{2F_{1/2}\left( \frac{1}{\xi}\right)}\sum\limits_{k=1}^{\infty }(-1)^{k}\Phi (z)\cos \left( 2\pi k\nu -\frac{\pi}{4}\right),  \nonumber
\end{eqnarray}
where $\Phi (z)=\frac{z\coth z-1}{z^{3/2}\cdot \sinh z}$ is a form-factor, $z=2\pi
^{2}\xi \nu k\sim kT/\hbar \omega _{c}$ is the dimensionless parameter which roughly corresponds to
quantum limit criteria at $k=1$. Then, $N_{0}=n_{0}\frac{3}{2}\xi ^{3/2}F_{1/2}(1/\xi )$ and
$S_{0}=N_{0}k\left( \frac{5F_{3/2}(1/\xi )}{3F_{1/2}(1/\xi
)}-\frac{1}{\xi }\right) $ are the zero-field electron gas density and entropy
respectively, $F_{n}(y)$ is the Fermi integral and $n_{0}=\frac{(2m\mu
)^{3/2}}{3\pi ^{2}\hbar ^{3}}$ is the density of strongly
degenerate 3D electron gas at $T=0$.

In Fig.\ref{fig:Fig.1} we plot the magnetic field dependencies of the Hall resistivity $\rho _{yx}(\nu
^{-1})$ and transverse magnetoresistivity $\rho (\nu ^{-1})$ given by
Eqs.(\ref{resistivity}-\ref{Lifshitz}). Both curves are
scaled in units of $\rho _{c}=\frac{m\mu }{n_{0}e^{2}\hbar }=
\frac{h}{e^{2}}\frac{3\pi }{4k_{F}}$, where $k_{F}=\sqrt{2m\mu
}/\hbar $ is the Fermi vector. Then, the longitudinal resistivity $\rho
_{zz}(\nu ^{-1})$ specified by Eq.(\ref{resistivity}) is scaled in
units of zero-field resistivity $\rho _{zz}(0)=\frac{m}{%
n_{0}e^{2}\tau }=\frac{2\rho _{c}}{k_{F}l}$, where $l=$ $\hbar
k_{F}\tau /m$ is the mean free path. The overall behavior of the resistivity
curves shown in Fig.\ref{fig:Fig.1} can be easily understood in terms of the
3D energy spectrum, and, thus the related density of states. Indeed, apart from the
discrete Landau levels, the 3D energy spectrum specified by Eq.(\ref{3D_spectrum})
contains the continuous component $\varepsilon _{\|}$. At
fixed $\mu $ the number of occupied states, and, hence $\rho
_{yx},\rho _{zz} \sim N^{-1}$ are smooth functions of the magnetic field.
Then, the density of states is continuous, hence provides the nonzero transverse
magnetoresistivity $\rho $. It is worthwhile to mention that in 2D electron gas case
the energy spectrum is purely discrete, thus leads to
quantized Hall resistivity $\rho _{yx}$, with the longitudinal resistivity
$\rho $ vanishing within the Hall plateaus[14].

We now make an attempt to compare our results with experimental
observations. As expected, the longitudinal, transverse and the Hall
components of magnetoresistivity exhibit familiar 1/H-period in-phase oscillations[20]
associated with discrete LL energy spectrum. However, in contrast to our predictions the behavior of the
longitudinal and the transverse magnetoresistivity components[20]
becomes uncorrelated at high fields when the only few LLs are filled.
The possible reason for the above discrepancy could be, for example, the constant momentum relaxation
time, i.e. $\tau \neq \tau(B)$, adopted in our simple model. For typical n-InSb sample $
n_{0}=1.2\cdot 10^{16}$cm$^{-3}$ used in Ref.[20] we obtain
$\mu =170$K and, then $\rho _{c}=0.1$Om$\cdot
$cm. At filling $\nu=3/2$ our zero spin-split approach provides(see Fig.\ref{fig:Fig.1}) the transverse
magnetoresistivity amplitude $\rho \simeq 0.002$\ Om$\cdot $cm at
liquid helium temperature $\xi =0.025$. This value is
consistent with that $\sim 0.015$\ Om$\cdot $cm observed experimentally[20]
at $B=2.3$T.

\section{Conclusions}
\label{Conclusions}
In conclusion, we have calculated the longitudinal, transverse and the Hall
magnetoresistivities of a 3D solid placed in quantizing magnetic taking into
account the contribution caused by combined action of Peltier and Seebeck
thermoelectric effects. The transport coefficients demonstrate familiar 1/H-period
oscillations being universal functions of the filling factor. The amplitude of the
transverse resistivity is consistent with that observed in experiment.

\section{Appendix}
\label{Appendix}
Let us examine the charge relaxation and screening for 3D
solids placed in magnetic field. These effects can be
described by continuity equation
\begin{equation}
\frac{\partial \rho _{e}}{\partial t}+div{\bf j}=0.
 \label{Maxwell}
\end{equation}
In quantizing magnetic field the non-dissipative transverse current
specified by Eq.(\ref{current+heat}) can be re-written[21]
as it follows ${\bf j}=-\frac{cN}{B^{2}}$rot$(\zeta {\bf B})$. Thus, the
charge relaxation is absent since div${\bf j}\equiv 0$, therefore
$\partial \rho _{e}/\partial t=0$. The retardation effects, if accounted, could result
in slow charge relaxation $\sim {\bf (}\sigma _{yx}/c)^{2}$ analogous
to that in 2D electron gas case[22]. We argue that the transverse length
of the Debye screening could be comparable with the macroscopic( i.e. sample) length
scale. We conclude that in quantizing magnetic field the electron
plasma neutrality may be violated in 3D sample bulk.

\end{document}